\documentclass[apj]{emulateapj}

\usepackage{amsmath,amstext,amssymb}
\usepackage{graphicx}
\usepackage{hyperref}
\usepackage[figure,figure*]{hypcap}
\usepackage{enumerate}
\usepackage{multirow}
\usepackage{amsfonts}
\usepackage{amssymb} 
\usepackage{bm}
\usepackage{natbib}
\usepackage{courier}
\usepackage{xcolor}

\mathchardef\mhyphen="2D

\shorttitle{The Evolution of AGN Activity in BCGs}
\shortauthors{Somboonpanyakul et al.}

\altaffiltext{\MIT}{Kavli Institute for Astrophysics and Space Research, Massachusetts Institute of Technology, 77 Massachusetts Avenue, Cambridge, MA 02139, USA}
\altaffiltext{\KIPAC}{Kavli Institute for Particle Astrophysics \& Cosmology, P. O. Box 2450, Stanford University, Stanford, CA 94305, USA}
\altaffiltext{\ASU}{School of Earth and Space Exploration, Arizona State University, Tempe, AZ, 85287, USA}
\altaffiltext{\LIABrazil}{Laborat\'orio Interinstitucional de e-Astronomia - LIneA, Rua Gal. Jos\'e Cristino 77, Rio de Janeiro, RJ - 20921-400, Brazil}
\altaffiltext{\Fermi}{Fermi National Accelerator Laboratory, P. O. Box 500, Batavia, IL 60510, USA}
\altaffiltext{\IFTBrazil}{Instituto de F\'{i}sica Te\'orica, Universidade Estadual Paulista, S\~ao Paulo, Brazil}
\altaffiltext{\ICPortsmouth}{Institute of Cosmology and Gravitation, University of Portsmouth, Portsmouth, PO1 3FX, UK}
\altaffiltext{\UCin}{Department of Physics, University of Cincinnati, Cincinnati, OH 45221, USA}
\altaffiltext{\CNRS}{CNRS, UMR 7095, Institut d'Astrophysique de Paris, F-75014, Paris, France}
\altaffiltext{\Sorbonne}{Sorbonne Universit\'es, UPMC Univ Paris 06, UMR 7095, Institut d'Astrophysique de Paris, F-75014, Paris, France}
\altaffiltext{\Sussex}{Department of Physics and Astronomy, Pevensey Building, University of Sussex, Brighton, BN1 9QH, UK}
\altaffiltext{\ULondon}{Department of Physics \& Astronomy, University College London, Gower Street, London, WC1E 6BT, UK}
\altaffiltext{\UChicago}{Department of Astronomy and Astrophysics, University of Chicago, 5640 South Ellis Avenue, Chicago, IL 60637}
\altaffiltext{\SLAC}{SLAC National Accelerator Laboratory, Menlo Park, CA 94025, USA}
\altaffiltext{\UPortsmouth}{Institute of Cosmology and Gravitation, University of Portsmouth, PO1 2UP, UK}
\altaffiltext{\IACSpain}{Instituto de Astrofisica de Canarias, E-38205 La Laguna, Tenerife, Spain}
\altaffiltext{\ULaLaguna}{Universidad de La Laguna, Dpto. Astrofísica, E-38206 La Laguna, Tenerife, Spain}
\altaffiltext{\CASUrbana}{Center for Astrophysical Surveys, National Center for Supercomputing Applications, 1205 West Clark St., Urbana, IL 61801, USA}
\altaffiltext{\UIUC}{Department of Astronomy, University of Illinois at Urbana-Champaign, 1002 W. Green Street, Urbana, IL 61801, USA}
\altaffiltext{\IFAE}{Institut de F\'{\i}sica d'Altes Energies (IFAE), The Barcelona Institute of Science and Technology, Campus UAB, 08193 Bellaterra (Barcelona) Spain}
\altaffiltext{\Trieste}{Astronomy Unit, Department of Physics, University of Trieste, via Tiepolo 11, I-34131 Trieste, Italy}
\altaffiltext{\INAF}{Osservatorio Astronomico di Trieste, via G. B. Tiepolo 11, I-34143 Trieste, Italy}
\altaffiltext{\IFPU}{Institute for Fundamental Physics of the Universe, Via Beirut 2, 34014 Trieste, Italy}
\altaffiltext{\ONRuaGal}{IObservat\'orio Nacional, Rua Gal. Jos\'e Cristino 77, Rio de Janeiro, RJ - 20921-400, Brazil}
\altaffiltext{\UMichigan}{Department of Physics, University of Michigan, Ann Arbor, MI 48109, USA}
\altaffiltext{\CIEMAT}{Centro de Investigaciones Energ\'eticas, Medioambientales y Tecnol\'ogicas (CIEMAT), Madrid, Spain}
\altaffiltext{\JPL}{Jet Propulsion Laboratory, California Institute of Technology, 4800 Oak Grove Drive, M/S 169-327, Pasadena, CA 91109, USA}
\altaffiltext{\SantaCruz}{Santa Cruz Institute for Particle Physics, Santa Cruz, CA 95064, USA}
\altaffiltext{\UMichiganAstro}{Department of Astronomy, University of Michigan, Ann Arbor, MI 48109, USA}
\altaffiltext{\UOslo}{Institute of Theoretical Astrophysics, University of Oslo. P.O. Box 1029 Blindern, NO-0315 Oslo, Norway}
\altaffiltext{\UMissouri}{Department of Physics and Astronomy, University of Missouri--Kansas City, 5110 Rockhill Road, Kansas City, MO 64110, USA}
\altaffiltext{\CSIC}{Instituto de Fisica Teorica UAM/CSIC, Universidad Autonoma de Madrid, 28049 Madrid, Spain}
\altaffiltext{\IEEC}{Institut d'Estudis Espacials de Catalunya (IEEC), 08034 Barcelona, Spain}
\altaffiltext{\ICE}{Institute of Space Sciences (ICE, CSIC), Campus UAB, Carrer de Can Magrans, s/n, 08193 Barcelona, Spain}
\altaffiltext{\UFlorida}{Department of Astronomy, University of Florida, Gainesville, FL 32611, USA}
\altaffiltext{\LMU}{Faculty of Physics, Ludwig-Maximilians-Universit\"at, Scheinerstr. 1, 81679 Munich, Germany}
\altaffiltext{\CSIRO}{CSIRO Astronomy \& Space Science, PO Box 1130, Bentley, WA 6102, Australia}
\altaffiltext{\UQueensland}{School of Mathematics and Physics, University of Queensland, Brisbane, QLD 4072, Australia}
\altaffiltext{\OhioState}{Department of Astronomy, The Ohio State University, Columbus, Ohio 43210, USA}
\altaffiltext{\OhioStateCos}{Center of Cosmology and Astro-Particle Physics, The Ohio State University, Columbus, Ohio, 43210, USA}
\altaffiltext{\CfA}{Center for Astrophysics $\vert$ Harvard \& Smithsonian, 60 Garden Street, Cambridge, MA 02138, USA}
\altaffiltext{\KICP}{Kavli Institute for Cosmological Physics, University of Chicago, 5640 South Ellis Avenue, Chicago, IL 60637}
\altaffiltext{\MacquarieU}{Australian Astronomical Optics, Macquarie University, North Ryde, NSW 2113, Australia}
\altaffiltext{\Lowell}{Lowell Observatory, 1400 Mars Hill Rd, Flagstaff, AZ 86001, USA}
\altaffiltext{\USaoPaulo}{Departamento de F\'isica Matem\'atica, Instituto de F\'isica, Universidade de S\~ao Paulo, CP 66318, S\~ao Paulo, SP, 05314-970, Brazil}
\altaffiltext{\TexasAM}{George P. and Cynthia Woods Mitchell Institute for Fundamental Physics and Astronomy, and Department of Physics and Astronomy, Texas A\&M University, College Station, TX 77843, USA}
\altaffiltext{\Radcliffe}{Radcliffe Institute for Advanced Study, Harvard University, Cambridge, MA 02138, USA}
\altaffiltext{\PrincetonU}{Department of Astrophysical Sciences, Princeton University, Peyton Hall, Princeton, NJ 08544, USA}
\altaffiltext{\ICatalana}{Instituci\'o Catalana de Recerca i Estudis Avan\c{c}ats, E-08010 Barcelona, Spain}
\altaffiltext{\MPE}{Max Planck Institute for Extraterrestrial Physics, Giessenbachstrasse, 85748 Garching, Germany}
\altaffiltext{\UMadison}{Physics Department, 2320 Chamberlin Hall, University of Wisconsin-Madison, 1150 University Avenue Madison, WI 53706-1390}
\altaffiltext{\INFN}{National Institute for Nuclear Physics, Via Valerio 2, I-34127 Trieste, Italy}
\altaffiltext{\INAFB}{Osservatorio Astronomico di Brera, Via Brera 28, 20121 Milano, Italy}
\altaffiltext{\USouthampton}{School of Physics and Astronomy, University of Southampton, Southampton, SO17 1BJ, UK}
\altaffiltext{\OakRidge}{Computer Science and Mathematics Division, Oak Ridge National Laboratory, Oak Ridge, TN 37831}
\altaffiltext{\Stanford}{Department of Physics, Stanford University, 382 Via Pueblo Mall, Stanford, CA 94305, USA}

\def\MIT{1}
\def\KIPAC{2}
\def\ASU{3}
\def\LIABrazil{4}
\def\Fermi{5}
\def\IFTBrazil{6}
\def\ICPortsmouth{7}
\def\UCin{8}
\def\CNRS{9}
\def\Sorbonne{10}
\def\Sussex{11}
\def\ULondon{12}
\def\UChicago{13}
\def\SLAC{14}
\def\UPortsmouth{15}
\def\IACSpain{16}
\def\ULaLaguna{17}
\def\CASUrbana{18}
\def\UIUC{19}
\def\IFAE{20}
\def\Trieste{21}
\def\INAF{22}
\def\IFPU{23}
\def\ONRuaGal{24}
\def\UMichigan{25}
\def\CIEMAT{26}
\def\JPL{27}
\def\SantaCruz{28}
\def\UMichiganAstro{29}
\def\UOslo{30}
\def\UMissouri{31}
\def\CSIC{32}
\def\IEEC{33}
\def\ICE{34}
\def\UFlorida{35}
\def\LMU{36}
\def\CSIRO{37}
\def\UQueensland{38}
\def\OhioState{39}
\def\OhioStateCos{40}
\def\CfA{41}
\def\KICP{42}
\def\MacquarieU{43}
\def\Lowell{44}
\def\USaoPaulo{45}
\def\TexasAM{46}
\def\Radcliffe{47}
\def\PrincetonU{48}
\def\ICatalana{49}
\def\MPE{50}
\def\UMadison{51}
\def\INFN{52}
\def\INAFB{53}
\def\USouthampton{54}
\def\OakRidge{55}
\def\Stanford{56}

\begin{document}

\title{The Evolution of AGN Activity in Brightest Cluster Galaxies}

\author{
T.~Somboonpanyakul\altaffilmark{\MIT,\KIPAC},
M.~McDonald\altaffilmark{\MIT},
A.~Noble\altaffilmark{\ASU},
M.~Aguena\altaffilmark{\LIABrazil},
S.~Allam\altaffilmark{\Fermi},
A.~Amon\altaffilmark{\KIPAC},
F.~Andrade-Oliveira\altaffilmark{\LIABrazil,\IFTBrazil},
D.~Bacon\altaffilmark{\ICPortsmouth},
M.~B.~Bayliss\altaffilmark{\UCin},
E.~Bertin\altaffilmark{\CNRS,\Sorbonne},
S.~Bhargava\altaffilmark{\Sussex},
D.~Brooks\altaffilmark{\ULondon},
E.~Buckley-Geer\altaffilmark{\Fermi,\UChicago},
D.~L.~Burke\altaffilmark{\KIPAC,\SLAC},
M.~Calzadilla\altaffilmark{\MIT},
R.~Canning\altaffilmark{\UPortsmouth},
A.~Carnero~Rosell\altaffilmark{\LIABrazil,\IACSpain,\ULaLaguna},
M.~Carrasco~Kind\altaffilmark{\CASUrbana,\UIUC},
J.~Carretero\altaffilmark{\IFAE},
M.~Costanzi\altaffilmark{\Trieste,\INAF,\IFPU},
L.~N.~da Costa\altaffilmark{\LIABrazil,\ONRuaGal},
M.~E.~S.~Pereira\altaffilmark{\UMichigan},
J.~De~Vicente\altaffilmark{\CIEMAT},
P.~Doel\altaffilmark{\ULondon},
P.~Eisenhardt\altaffilmark{\JPL},
S.~Everett\altaffilmark{\SantaCruz},
A.~E.~Evrard\altaffilmark{\UMichigan,\UMichiganAstro},
I.~Ferrero\altaffilmark{\UOslo},
B.~Flaugher\altaffilmark{\Fermi},
B.~Floyd\altaffilmark{\UMissouri},
J.~Garc\'ia-Bellido\altaffilmark{\CSIC},
E.~Gaztanaga\altaffilmark{\IEEC,\ICE},
D.~W.~Gerdes\altaffilmark{\UMichigan,\UMichiganAstro},
A.~Gonzalez\altaffilmark{\UFlorida},
D.~Gruen\altaffilmark{\LMU}, 
R.~A.~Gruendl\altaffilmark{\CASUrbana,\UIUC}, 
J.~Gschwend\altaffilmark{\LIABrazil,\ONRuaGal},
N.~Gupta\altaffilmark{\CSIRO},
G.~Gutierrez\altaffilmark{\Fermi},
S.~R.~Hinton\altaffilmark{\UQueensland},
D.~L.~Hollowood\altaffilmark{\SantaCruz},
K.~Honscheid\altaffilmark{\OhioState,\OhioStateCos},
B.~Hoyle\altaffilmark{\LMU}, 
D.~J.~James\altaffilmark{\CfA}, 
T.~Jeltema\altaffilmark{\SantaCruz},
G.~Khullar\altaffilmark{\UChicago,\KICP},
K.~J.~Kim\altaffilmark{\UCin},
M.~Klein\altaffilmark{\LMU},
K.~Kuehn\altaffilmark{\MacquarieU,\Lowell},
M.~Lima\altaffilmark{\LIABrazil,\USaoPaulo},
M.~A.~G.~Maia\altaffilmark{\LIABrazil,\ONRuaGal},
J.~L.~Marshall\altaffilmark{\TexasAM},
P.~Martini\altaffilmark{\OhioState,\OhioStateCos,\Radcliffe},
P.~Melchior\altaffilmark{\PrincetonU},
F.~Menanteau\altaffilmark{\CASUrbana,\UIUC},
R.~Miquel\altaffilmark{\IFAE,\ICatalana},
J.~J.~Mohr\altaffilmark{\LMU,\MPE},
R.~Morgan\altaffilmark{\UMadison},
R.~L.~C.~Ogando\altaffilmark{\LIABrazil,\ONRuaGal},
A.~Palmese\altaffilmark{\Fermi,\KICP},
F.~Paz-Chinch\'{o}n\altaffilmark{\CASUrbana,\UIUC},
A.~Pieres\altaffilmark{\LIABrazil,\ONRuaGal},
A.~A.~Plazas~Malag\'on\altaffilmark{\PrincetonU},
K.~Reil\altaffilmark{\SLAC},
A.~K.~Romer\altaffilmark{\Sussex},
F.~Ruppin\altaffilmark{\MIT},
E.~Sanchez\altaffilmark{\CIEMAT},
A.~Saro\altaffilmark{\Trieste,\INAF,\IFPU,\INFN},
V.~Scarpine\altaffilmark{\Fermi},
M.~Schubnell\altaffilmark{\UMichigan},
S.~Serrano\altaffilmark{\IEEC,\ICE},
I.~Sevilla-Noarbe\altaffilmark{\CIEMAT},
P.~Singh\altaffilmark{\INAF,\IFPU},
M.~Smith\altaffilmark{\USouthampton},
M.~Soares-Santos\altaffilmark{\UMichigan},
V.~Strazzullo\altaffilmark{\INAF,\IFPU,\INAFB},
E.~Suchyta\altaffilmark{\OakRidge},
M.~E.~C.~Swanson\altaffilmark{\CASUrbana},
G.~Tarle\altaffilmark{\UMichigan},
C.~To\altaffilmark{\KIPAC,\SLAC,\Stanford},
D.~L.~Tucker\altaffilmark{\Fermi},
and R.~D.~Wilkinson\altaffilmark{\Sussex}
}

\begin{abstract}
We present the results of an analysis of Wide-field Infrared Survey Explorer (WISE) observations of the full 2500 $\rm{deg}^{2}$ South Pole Telescope (SPT)-Sunyaev–Zel’dovich cluster sample. We describe a process for identifying active galactic nuclei (AGN) in brightest cluster galaxies (BCGs) based on WISE mid-IR color and redshift. Applying this technique to the BCGs of the SPT-SZ sample, we calculate the AGN-hosting BCG fraction, which is defined as the fraction of BCGs hosting bright central AGNs over all possible BCGs. Assuming an evolving single-burst stellar population model, we find statistically significant evidence ($> 99.9\%$) for a mid-IR excess at high redshift compared to low redshift, suggesting that the fraction of AGN-hosting BCGs increases with redshift over the range of $0 < z < 1.3$. The best-fit redshift trend of the AGN-hosting BCG fraction has the form $(1+z)^{4.1\pm1.0}$. These results are consistent with previous studies in galaxy clusters as well as as in field galaxies. One way to explain this result is that member galaxies at high redshift tend to have more cold gas. While BCGs in nearby galaxy clusters grow mostly by dry mergers with cluster members, leading to no increase in AGN activity, BCGs at high redshift could primarily merge with gas-rich mergers, providing fuel for feeding AGNs. If this observed increase in AGN activity is linked to gas-rich mergers, rather than ICM cooling, we would expect to see an increase in scatter in the $P_{\rm cav}$ versus $L_{\rm cool}$ relation at $z>1$. Last, this work confirms that the runaway cooling phase, as predicted by the classical cooling-flow model, in the Phoenix cluster is extremely rare and most BCGs have low (relative to Eddington) black hole accretion rates. 
\end{abstract}


\section{Introduction}
Galaxy clusters are the most massive gravitationally bound and collapsed objects in the universe~\citep{Voit2005b}. Because of their extremely deep potential wells, the temperature of the intracluster medium (ICM) is high enough to radiate X-rays. The central parts of clusters, which have the densest X-ray-emitting gas, often have their cooling times shorter than the Hubble time, implying that the hot X-ray gas should have had enough time to cool and form large inward flows of cooling material, known as cooling flows~\citep{Sarazin1986, Fabian1994}. However, multi-wavelength observations have only seen a fraction of the massive cooling flows that are expected from standard cooling models~\citep[e.g.,][]{ODea2008, Donahue2015, McDonald2018}. This is referred to as ``the cooling-flow problem", and active galactic nuclei (AGN) feedback is thought to be responsible for preventing the hot gas from cooling by propagating energy from the supermassive black hole (SMBH) to the ICM. The two primary modes of AGN feedback are the kinetic mode, with relativistic jets pushing aside the hot gas and creating cavities, and the quasar mode, with the radiation coming from the accretion disk~\citep[see reviews by][]{Fabian2012, McNamara2012}.

With the recent development of galaxy cluster surveys which utilize the Sunyaev-Zel'dovich (SZ) effect~\citep{Sunyaev1972}, such as the South Pole Telescope~\citep[SPT;][]{Carlstrom2011,Bleem2015,Bleem2020,Huang2020} and the Atacama Cosmology Telescope~\citep[ACT;][]{Hilton2018,Hilton2021}, the number of known high-z ($z>1$) clusters with good mass estimates has increased dramatically. This has enabled many studies of the evolution of AGN feedback in clusters over cosmic time~\citep{McDonald2013, McDonald2017,Gupta2020}. However, the evolution of AGN feedback in galaxy clusters with redshift remains poorly understood. In particular, \citet{Hlavacek-Larrondo2015} found no evidence for evolution in jetted power generated by AGN feedback from X-ray cavities over the past 7 Gyr ($z=0.8$). An earlier study by~\citet{Hlavacek-Larrondo2013} suggested that the fraction of brightest cluster galaxies (BCGs) with X-ray bright nuclei is decreasing with time (or increasing with redshift), suggesting a strong evolution in radiative mode feedback. In contrast, a recent study looking for nuclear BCG X-ray emission in Chandra archival data instead found no evidence for evolution between two redshift bins ($\langle z \rangle\sim0.25$ and $\langle z \rangle\sim0.65$)~\citep{Yang2018}. The disagreement between various studies about the evolution of AGN feedback restricts our ability to fully understand this issue.

In this work, we calculate the AGN-hosting BCG fraction by identifying BCGs in the 2500 deg$^{2}$ SPT-SZ cluster samples~\citep{Bleem2015} and classifying whether they are AGNs, based on mid-infrared data. The SZ cluster catalogs allow for an effectively mass-selected sample of clusters, making it possible to study the evolution of galaxy clusters over time. In addition, the SZ catalogs typically have less contamination, compared to optical/IR catalogs. The fraction of BCGs hosting luminous AGNs is an important indicator for AGN fueling processes, availability of cold clumpy gas in the centers of clusters, the duration and duty cycle of the AGNs, and how BCGs and the host clusters grow and evolve together. This is because additional physical mechanisms are often required to explain the transport of the cold gas, which serves as the primary fuel source for the central black holes. The fact that we find a relative absence of AGNs in the centers of clusters has led us to study many physical processes, including ram-pressure stripping~\citep{Gunn1972}, tidal effects from the cluster gravitational potential~\citep{Merritt1983}, and the lack of new infall of cold gas~\citep{Larson1980}. Similarly, understanding the evolution of AGN activities in BCGs will help us understand the evolution mechanism of galaxy clusters, and how the feedback might play a role in that.

Our goal for this paper is to study the redshift evolution of the AGN-hosting BCG fraction up to $z=1.3$ to understand the fueling processes in the centers of clusters, determine when AGN feedback is fully established, and identify whether there are any more extreme AGN-hosting BCGs in the sample, similar to the Phoenix cluster. The paper is organized as follows. In Section~\ref{sec::data} and Section~\ref{sec::method} we summarize the data and additional information used in this paper. The results and their implications are presented in Section~\ref{sec::result} and Section~\ref{sec::discussion}, respectively. We conclude our work in Section~\ref{sec::conclusion}. We assume $H_0 = 70\,\rm{km\,s^{-1}\,Mpc^{-1}}$, $\Omega_{\rm m}= 0.3$ and $\Omega_{\rm \lambda} = 0.7$. All errors are $1\sigma$ unless noted otherwise.

\section{Data} \label{sec::data}

\subsection{The SPT-SZ 2500 deg$^{2}$ Cluster Sample} \label{sec::spt}
We use the full 2500 deg$^{2}$ SPT-SZ cluster sample from~\citet{Bleem2015} with the improvement in the cluster redshift estimates from~\citet{Bocquet2019} by incorporating new spectroscopic and improved photometric measurements~\citep{Bayliss2016,Khullar2019}. The survey spans a contiguous 2500 deg$^{2}$ area within a boundary of RA = 20h--7h, and Dec. = -65$^{\circ}$--\,-40$^{\circ}$. Once we limit the redshift range to $0<z<1.3$, the total number of clusters in our sample is 475. 

\subsection{Position of BCGs} \label{sec::bcg}
Given the diversity of BCG colors, morphologies, and assembly state as a function of redshift, typical identification algorithms may be biased when they select BCGs based on single-band fluxes.  We have instead developed a novel BCG identification pipeline that utilizes the full probability distribution of redshift and stellar mass for every object within 500 projected kiloparsec of the SZ cluster center to assign BCG likelihoods.  Photometry is provided by the Dark Energy Survey (Year 3) catalogs~\citep{Jarvis2021,Sevilla-Noarbe2021}, cross-correlated with unWISE~\citep{Lang2014}, which is a combination of WISE and NEOWISE images. Various cuts and flags are utilized to avoid stars, and objects with poor photometric measurements. WISE is an infrared-wavelength space satellite with four IR filters, including $W1$ ($\lambda_{cen} = 3.6$ $\mu m$), $W2$ (4.3 $\mu m$), $W3$ (12 $\mu m$), and $W4$ (22 $\mu m$)~\citep{Wright2010}. The satellite operated for two years with cryogen until 2011, before being reactivated and resuming operations as NEOWISE in 2013, and has continued to observe ever since~\citep{Mainzer2014}. 

\begin{figure*}[ht]
	\begin{center}
		\includegraphics[width=2\columnwidth]{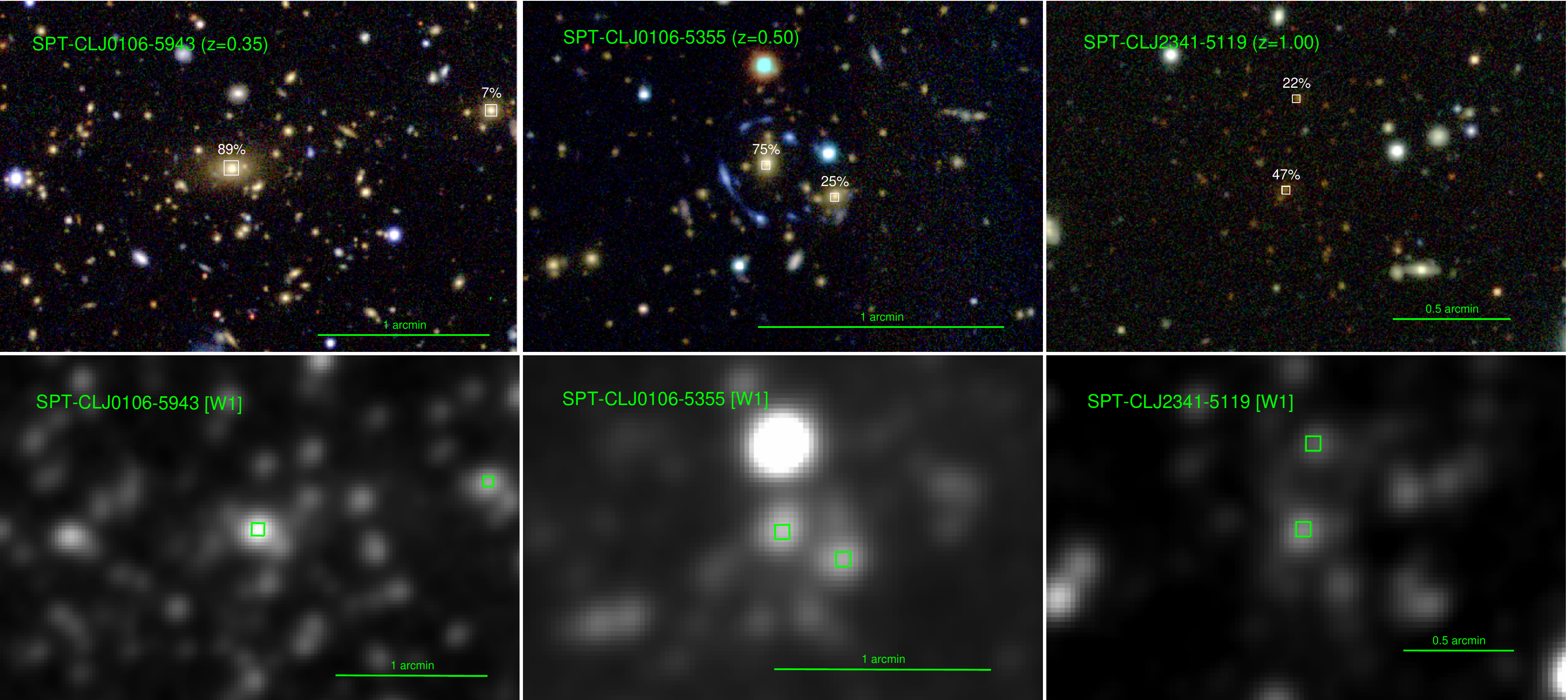}
		\caption{The top three panels show DES $gri$ optical images of three SPT galaxy clusters, including SPT-CLJ0106-5943, SPT-CLJ0106-5355, and SPT-CLJ2341-5119. The white boxes show the location of the two highest-probability BCG candidates for each cluster while the white numbers show the probability (in percentage) for each object to be a true BCG. The three examples are ranging from $z=0.35$ to $z=1.00$, demonstrating the ability for this method to find BCG candidates up to high redshift. The bottom three panels show the corresponding WISE images from W1 channel for the three SPT clusters. The green boxes show the same location of the two highest-probability BCG candidates, suggesting that the BCG candidates are detectable in mid-IR.}
		\label{fig::bcg}
	\end{center}
\end{figure*}

Probability distributions of photometric redshift and stellar mass for each source are estimated with EAZY~\citep{Brammer2008} and FAST~\citep{Kriek2009}, respectively. We then randomly sample from each distribution to find the most massive cluster galaxy at the cluster redshift within each field, iterating this process $10^5$ times to build up a BCG likelihood for each galaxy.  In this way, all galaxies are assigned a value between 0 to 100\% probability of being the BCG within each cluster. Full details on the pipeline, along with the BCG catalog, will be provided in Noble et al. (in preparation). The top three panels of Fig.~\ref{fig::bcg} show optical images of example SPT galaxy clusters with identified BCGs in white squares. This demonstrates that the algorithm selects likely BCGs that match the galaxies that typical/traditional visual BCG identification methods would select over a wide range of redshift.

\subsection{Data for AGN Selection} \label{sec::wise}
Most photometric techniques for identifying AGNs are severely biased toward unobscured (type 1) AGNs since their nuclear emissions dominate over host galaxies, making these AGNs easily identifiable. This implies that most obscured (type 2) AGNs are often underrepresented in most studies. The most promising techniques for identifying both obscured and unobscured AGNs include radio, hard X-ray, and mid-infrared selections. However, not all AGNs are radio loud~\citep[e.g.,][]{Stern2000} and the current hard X-ray satellites remain limited in their sensitivity and field of view. This leaves mid-infrared selection as a popular technique to quickly identify large AGN populations (obscured and unobscured). The idea of mid-infrared selection is to separate between the power-law AGN spectrum and the blackbody stellar spectrum of galaxies, which peaks at rest frame $1.6\,\mu m$. The power-law spectra of the AGNs is due to the thermal emission from the warm-hot dust in the torus, which is heated by absorbing shorter wavelength photons from the accretion disk~\citep{Stern2012,Hickox2018}. This implies that the emission is not strongly suppressed by the dusty torus, unlike UV to near-IR wavelength, allowing this technique to detect more obscured AGNs. Additionally, with the first all-sky data release of Wide-field Infrared Survey Explorer~\citep[WISE;][]{Wright2010} in 2012, mid-infrared selection became one of the top methods of probing the AGN population over the entire sky without additional observations.

One drawback of the mid-IR selection technique is that the host galaxy is still bright at these wavelengths, limiting the detection of low-luminosity AGN which have to compete with a bright stellar continuum~\citep{Stern2012}. This means that AGNs selected by their mid-IR color tend to be brighter relative to the host galaxies than those selected by other techniques. For example, \citet{Assef2013} found that in the sample of relatively luminous AGNs ($L_{\rm AGN}/L_{\rm host}>0.5$), the luminosity of mid-IR AGNs tends to be greater than $\sim\!5\times10^{44}\,{\rm erg\,s^{-1}}$, taking into account the bolometric correction from~\citet{Singal2016}. Assuming the efficiency of turning accreting matter into energy $\epsilon_{acc}=0.1$ and a typical mass of the supermassive black hole $M_{\rm BH}\sim\!10^{8}$--$10^{9}\,M_{\odot}$~\citep{Russell2013}, the black hole mean accretion rate ($\dot{M}_{BH}=L_{\rm bol,nuc}/\epsilon_{acc}c^{2}$) of mid-IR selected AGNs should be greater than $4\times10^{-3}$--$4\times10^{-2}\,\dot{M}_{\rm Edd}$, where $\dot{M}_{\rm Edd}$ is the limiting Eddington accretion rate. This level of accretion is relatively high, compared to typical optical/radio AGNs, which have an accretion rate at around $10^{-6}$--$10^{-2}\,\dot{M}_{\rm Edd}$~\citep{McDonald2021}. This implies that the mid-IR technique will identify mainly a brighter and more massive AGN population. From now on, AGNs mentioned in this paper mean the mid-IR-selected AGN population. 

\subsubsection{Mid-IR Data from WISE}
Instead of using the main source catalog from WISE (AllWISE) that only includes the data obtained from 2010 to 2011, we make use of CatWISE to get the best photometry with available data. CatWISE is an updated all-sky infrared source catalog, which combines the 2010 and 2011 data from WISE with the 2013 through 2016 NEOWISE data~\citep{Eisenhardt2020}. The caveat is that the CatWISE catalog only includes 3.6 ($W1$) and 4.3 ($W2$) $\mu m$ data. In this work, we use the Preliminary CatWISE catalog to obtain the WISE color for each BCG, which has an advantage of including four times as many exposures as that used for the AllWISE catalog while using the same AllWISE software, making it more straightforward to make a comparison with previous works.

For every BCG identified (with probability $>5\%$) from Section~\ref{sec::bcg}, we search for mid-IR counterparts in the CatWISE catalog within a radius of 3$^{\prime\prime}$ from the identified BCG, since the typical FWHM for $W1$ and $W2$ are 6.08$^{\prime\prime}$ and 6.84$^{\prime\prime}$, respectively. Both $W1$ and $W2$ are converted from Vega to AB magnitudes using the correction from the Explanatory Supplement to the WISE Products\footnote{wise2.ipac.caltech.edu/docs/release/allsky/expsup/sec4\_4h.html}. The bottom three panels of Fig.~\ref{fig::bcg} shows $W1$ images of the three SPT clusters, showing that their BCG candidates can be detected with WISE data.

\subsubsection{CatWISE Color Correction}
\begin{figure}[ht]
	\begin{center}
		\includegraphics[width=1.0\columnwidth]{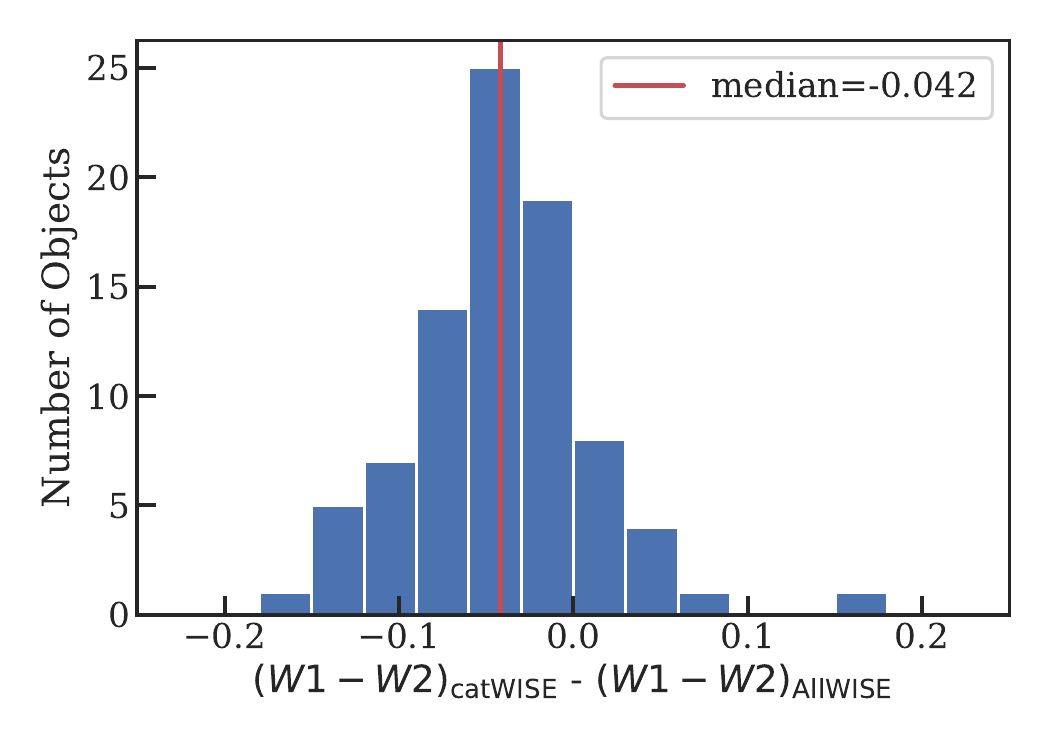}
		\caption{This figure shows a comparison of $W1-W2$ between AllWISE and CatWISE. The red line is the median of the difference at 0.042 mag.}
		\label{fig::catwise_offset}
	\end{center}
\end{figure}

We perform a comparison test between the AllWISE and CatWISE catalogs. The test is carried out by comparing the $(W1-W2)_{\rm{AB}}$ color of bright objects ($16.5<W1_{\rm{AB}}<18$) between the two catalogs within 10 arcmins of one field. A histogram of the color differences is plotted in Fig.~\ref{fig::catwise_offset}, showing the offset between the color from AllWISE and CatWISE to be around 0.042. Further investigation reveals that this is due to the gradual diminishing of the $W2$ throughput with time, leading to a bluer $(W1-W2)_{\rm{AB}}$ color, compared to AllWISE. We apply this correction of 0.042 mag to CatWISE $W1-W2$ colors. 

\section{Method} \label{sec::method}

\subsection{AGN Selection}\label{sec::agn_selection}
With the \textit{WISE} satellite, \citet{Stern2012} developed a well-known formula to quickly identify AGN candidates with a simple color criterion, $(W1-W2)_{\rm{Vega}}\ge0.8$ or $(W1-W2)_{\rm{AB}}\ge0.16$. One benefit of mid-IR selected samples is that both unobscured (type 1) and obscured (type 2) AGN can be identified~\citep{Lacy2004,Stern2005,Stern2012}. However, since the colors of galaxies drastically change over a large redshift range, this simple criterion is not accurate enough to characterize a large population of AGNs. To increase the number of AGNs we can identify, we develop a new color criterion which depends on redshifts of galaxies.

\subsubsection{EzGal Galaxy Color Model} \label{sec::ezgal}
\begin{figure}[ht]
	\begin{center}
		\includegraphics[width=1\columnwidth]{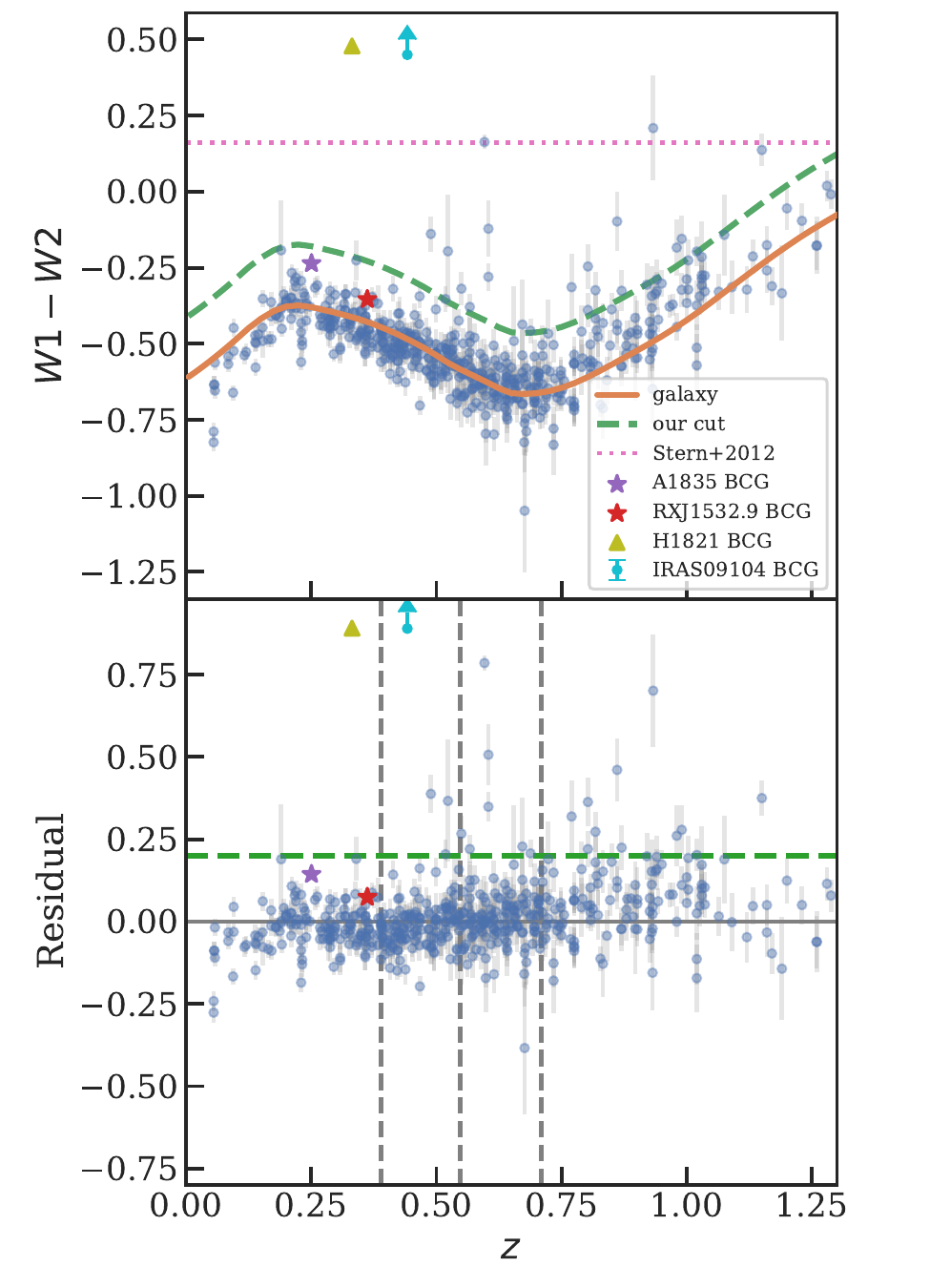}
		\caption{Top: $W1$-$W2$ color for each BCG candidate as a function of cluster redshift. The orange solid line shows the expected color as a function of redshift for our elliptical galaxy model using EzGal, as described in Section~\ref{sec::ezgal}. The green dashed line shows our criterion for selecting AGNs, which is derived from the orange line, and the pink dotted line shows the cut from the previous work by~\citet{Stern2012}. Bottom: the $W1$-$W2$ color difference between each BCG candidate and the expected color. The green dashed line shows our selection with the residual $> 0.2$. Every object with the residual greater than 0.2 is likely to be an AGN. The vertical gray dashed lines show the binning for the results in Fig.~\ref{fig::frac_agn}. The redshift bins are defined such that each bin contains roughly the same number of systems, making for uniform counting statistics across all redshifts. The two colored stars are known BCGs in galaxy clusters with large SFR, showing that our selection criterion does not select these starburst BCGs, while the two clusters with luminous AGNs (H 1821+643 and IRAS 09104+4109) are clearly above our criterion.}
		\label{fig::w1mw2}
	\end{center}
\end{figure}

To determine whether each BCG harbors an AGN, we calculate the expected color for typical elliptical galaxies using EzGal\footnote{www.baryons.org/ezgal/}. EzGal calculates the magnitude evolution as a function of redshift from evolving the spectral energy distribution (SED) models of a stellar population with time and projecting them through filters~\citep{Mancone2012}. This calculation takes into account both the stellar evolution of a galaxy as young stars evolves as well as the wavelength shift due to the distance of a galaxy. To find the model that best describes our overall sample, we perform a grid search between three stellar population model sets~\citep[i.e.,][]{Bruzual2003,Conroy2009,Maraston2005}, various formation redshifts, two different initial mass functions (IMF)~\citep[i.e.,][]{Salpeter1955,Chabrier2003}, star formation history as a single exponential decaying burst of star formation with an e-folding time parameter ($\tau$) between 0.1 and 10 Gyr, and the representative metallicity ($Z$) for our galaxy sample from 0.001 to 0.03. Ultimately, the best-fit model (based on the Chi-square test) is the \citet{Bruzual2003} stellar model with a formation redshift of ($z_f$) 3.5, the \citet{Salpeter1955} IMF, $\tau$ = 0.1 Gyr for star formation history, and the metallicity of 0.016. The orange solid line in the top panel of Fig.~\ref{fig::w1mw2} shows the expected $W1-W2$ color evolution, generated from the EzGal model with this particular set of parameters. The bottom panel shows the residual from the expected value of $W1$-$W2$ for each BCG. It demonstrates that the scatter is distributed around zero with a relatively weak redshift dependence, implying that we have successfully removed the continuum contribution. In this work, galaxies which are redder (the residual greater than 0.2) than typical elliptical galaxies based on the EzGal model are considered AGN candidates. For a range of threshold values from 0.2 to 0.4, the highest redshift bin has more AGN-hosting BCGs, compared to the lowest redshift bin, implying that an increase in the fraction of AGN-hosting BCGs is independent of this choice. To further test this notion, we perform the heteroscedasticity test, specifically the Breusch-Pagan test, on the bottom panel of Fig.~\ref{fig::w1mw2}, which shows whether the scatter of the infrared residual depends on redshift, regardless of the choice of a threshold value. The test results in P-value $= 0.0047$, meaning that the heteroscedasticity is present and the scatter of $W1-W2$ residuals depends on the redshift, implying that an increase in the fraction is feasible regardless of the threshold choice.

\begin{figure*}[ht]
	\begin{center}
		\includegraphics[width=1.95\columnwidth]{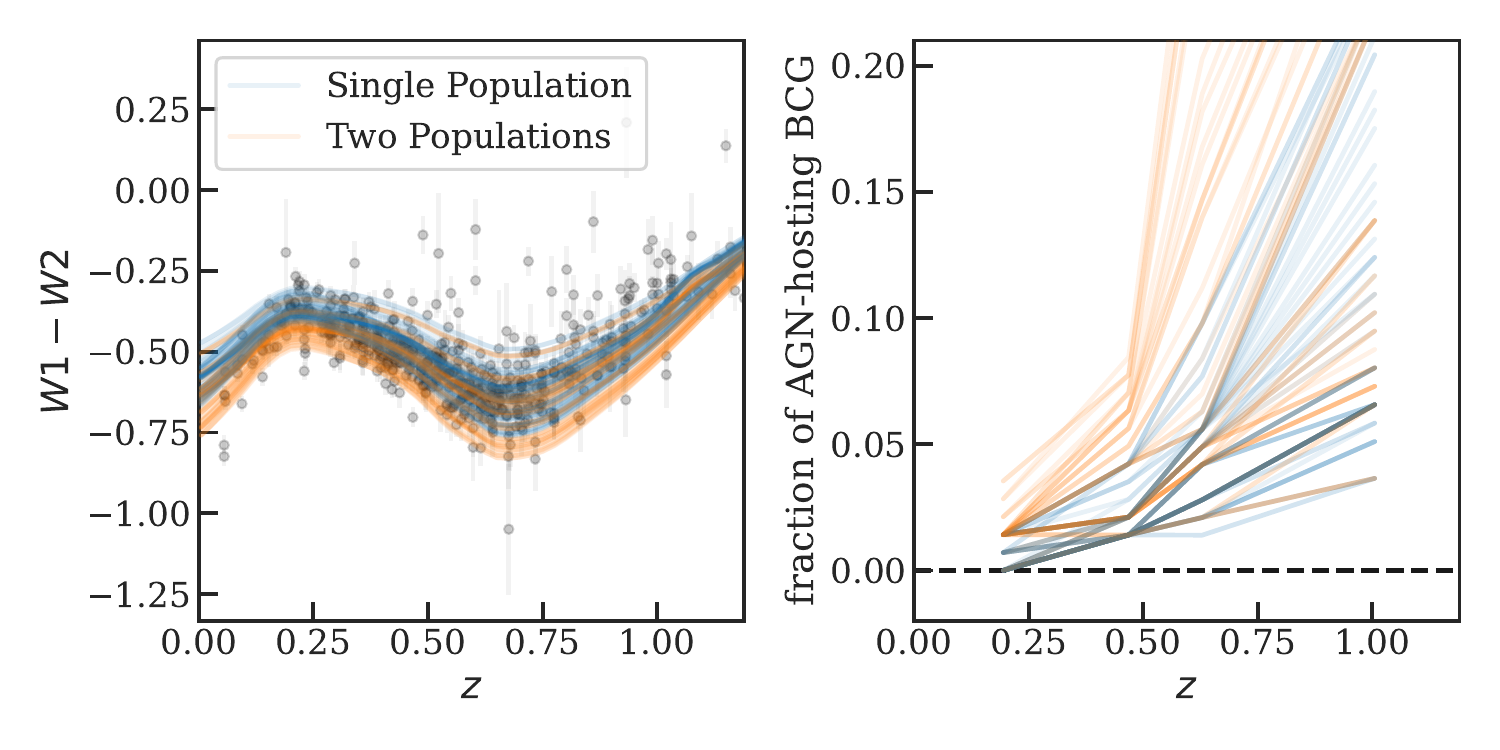}
		\caption{Left: W1-W2 colors (black dots) for each BCG candidate in this sample. The solid lines show the expected color as a function of redshift using EzGal models for both single-burst stellar population (blue) and the two-age stellar population models (orange). The single-burst models are plotted from a wide range of parameters, including two IMF~\citep{Salpeter1955,Chabrier2003}, three formation redshift ($z_f=1.5,2.5,3.5$), and a span of metallicity from 0.75 to 1.0 $Z_{\odot}$. To get the best fit, we perform a grid search between the IMF and the three formation redshifts before fitting the metallicity to minimize Chi-square. The old and young stellar models use the same parameter sets but include an additional `young' population, which is represented by a 50 Myr old stellar population at all observed redshifts. Right: The fraction of AGN-hosting BCGs as a function of redshift for all models. The figure demonstrates that the fraction of AGN-hosting BCGs increases with redshift regardless of our choice of stellar population model.} 
		\label{fig::many_model}
	\end{center}
\end{figure*}

One assumption that we apply in this section is that we only consider a single-burst stellar population model with a single formation redshift, star formation history, and metallicity. In Fig.~\ref{fig::many_model}, we consider both a single-burst stellar population model and a more complicated two-age stellar population (old and young) model with a wide range of parameters for both models. The two stellar population model keeps the same parameter sets from a single-burst model for the `old' population, while a `young' population is represented by a 50 Myr old stellar population at all redshifts. Even though these two models are likely not sufficient to describe our data, more sophisticated models would be unconstrained by the data that we have available. Based on these single-age and two-age models, we find no combination of formation time, metallicity, and IMF that can fully account for the observed evolution in the mid-IR excess, as shown in the right panel of Fig.~\ref{fig::many_model}. We propose that this mid-IR excess comes from a dusty torus, a signature of an actively accreting supermassive black hole. It is difficult to imagine other astronomical sources for this emission, since star formation typically yields significantly cooler dust temperatures, with the peak brightness around $\sim$100$\mu m$, instead of $\sim$1--10$\mu m$. On the other hand, it could also be that our current population models are not adequate to describe the data. We should keep this caveat in mind when we discuss the implication of our results. 

As a test to see how a starburst can affect the mid-IR color, we consider Abell 1835~\citep{Ehlert2011} and RX J1532.9+3021~\citep{Hlavacek-Larrondo2013b}, which are the most star forming BCGs known~\citep[SFR $\sim\!100\,M_{\odot}\,\rm{yr}^{-1}$][]{McDonald2018} that also lack evidence of a strong AGN. The two colored stars in Fig.~\ref{fig::w1mw2} demonstrate that even though a star-forming BCG would have boosted mid-IR emission due to dust, polycyclic aromatic hydrocarbon molecules (PAH), and molecular gas, the emission is not as strong as the power-law spectra of AGNs, and our selection does not include these two BCGs. On the other hand, the two clusters with the most luminous AGNs (H 1821+643~\citep{Russell2010} and IRAS 09104+4109~\citep{2012OSullivan}) are easily detected with our criterion.

\subsubsection{Spitzer Color Verification} \label{sec::spitzer}
Because the point-spread function (PSF) of the two WISE bands are not small (PSF$_{W1} = 6.^{\prime\prime}08$ and PSF$_{W2} = 6.^{\prime\prime}84$), we compare the results from \textit{WISE} mid-IR color with those from the \textit{Spitzer} Space Telescope. \textit{Spitzer} is an infrared telescope with the Infrared Array Camera~\citep[IRAC;][]{Fazio2004} as one of its main science instruments. IRAC is a four-channel imaging camera capable of taking simultaneous images at wavelengths of 3.6, 4.5, 5.8, and 8.0 $\mu m$. Thus, the channel 1 and 2 ([3.6] and [4.5]) on IRAC are roughly equivalent with $W1$ and $W2$ from \textit{WISE}, but with a benefit of having a much better PSF at $1.^{\prime\prime}95$ and $2.^{\prime\prime}02$, respectively~\citep{Fazio2004}. 

\begin{figure}[ht]
	\begin{center}
		\includegraphics[width=0.92\columnwidth]{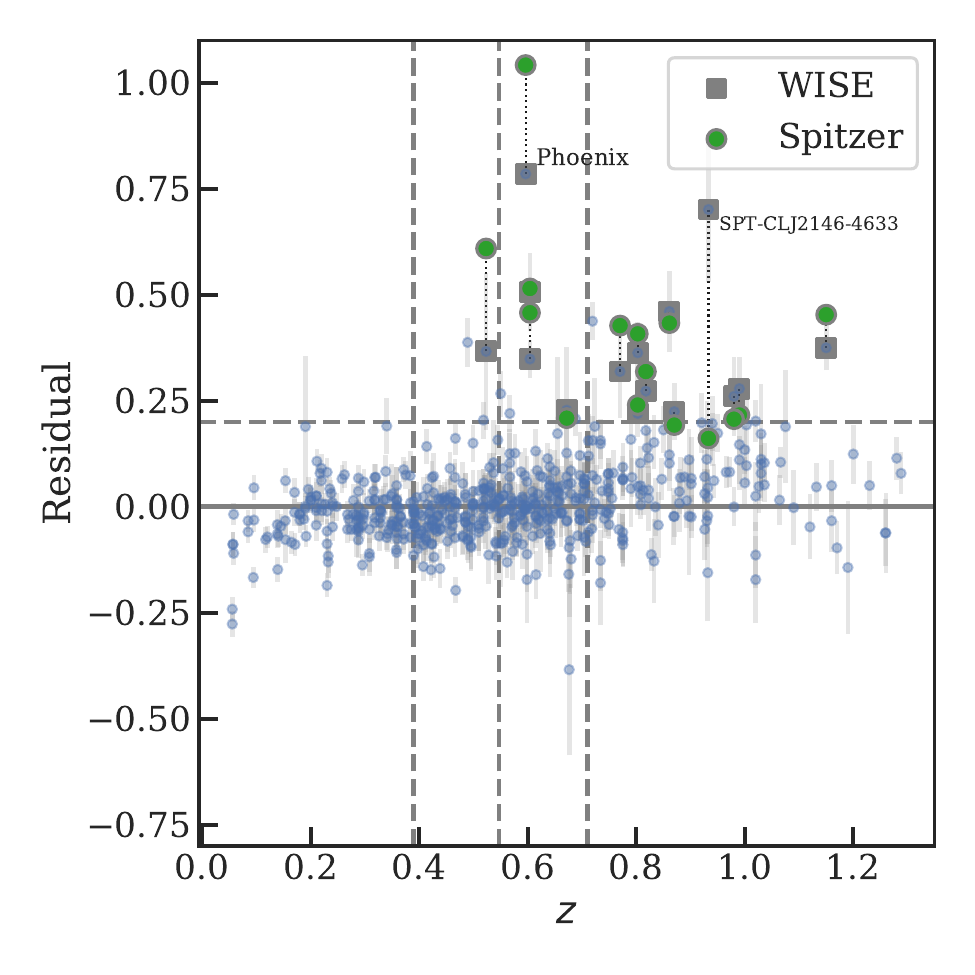}
		\caption{Residual plot similar to the bottom panel of Fig.~\ref{fig::w1mw2}. The blue points are the $W1$-$W2$ color difference between each BCG candidate and the expected color. The gray squares emphasize the $W1$-$W2$ color from WISE for the sample that has been observed by Spitzer, while the green circles show the color from Spitzer, demonstrating that most of the objects we classify as AGNs have a difference of Spitzer and WISE color to be either compatible (60\% of the sources have a difference to be within $\pm$0.07 mag) or Spitzer is slightly redder (33\% of the sources have the Spitzer color larger than the WISE one by $\sim$0.2 mag). This means the color and our results are not strongly impacted by WISE's larger PSF.}
		\label{fig::spitzer}
	\end{center}
\end{figure}

A certain fraction ($\sim$35\%, predominantly at $z>0.8$) of the SPT cluster sample has been observed with IRAC. For verification, we compare the [3.6]-[4.5] colors of our AGN-hosting BCG candidates with their $W1-W2$ colors. If the Spitzer color, which has a higher angular resolution, is bluer (smaller) than the WISE color, it shows that there is a contamination from nearby galaxies within the WISE aperture. On the other hand, if the Spitzer color is redder (larger), it implies that the object is even more likely to be an AGN. Fig.~\ref{fig::spitzer} shows the comparison between WISE's $W1-W2$ (in gray squares) and Spitzer's [3.6]-[4.5] color (in circles) for our AGN candidates. We find that most AGN candidates have a difference of Spitzer and WISE color to be either compatible (60\% of the sources have a difference to be within $\pm$0.07 mag which is roughly the mean of the WISE color uncertainty) or Spitzer is slightly redder (33\% of the sources have the Spitzer color larger than the WISE one by $\sim$0.2 mag). This suggests that most of our AGN candidates are likely to be real quasars. One clear exception is SPT-CL J2146-4633, which has a WISE color much redder than Spitzer. Further investigation shows that there is a point-like source near the location of the object, but not at the BCG location, meaning that the WISE color is probably contaminated by a nearby AGN while the Spitzer color is not. This object has been removed in the further analysis.

\subsection{Calculating the AGN fraction}
We compute the fraction of galaxy clusters with AGN-hosting BCGs based on the number of BCGs whose mid-IR colors are redder than expected, as described in Section~\ref{sec::ezgal}, in four redshift bins to study the redshift evolution. With the probability estimated in Section~\ref{sec::bcg}, we first include in Fig.~\ref{fig::w1mw2} all the BCGs which are identified with a probability that is higher than 20\%, meaning that some clusters will have more than one BCG candidate. The fraction of AGN-hosting BCGs are calculated over the total number of BCG candidates, instead of the total number of clusters in each bin. We will discuss these particular choices of calculating AGN-hosting BCG fraction in Section~\ref{sec::result}. The bins are defined from $z=$ 0--1.3 in such a way that each bin contains roughly the same number of BCG candidates ($\sim$140 BCGs, see Table~\ref{table::agn_number} for the exact number). This choice of binning yields uniform counting statistics across all redshifts. The uncertainties associated with the AGN fractions are estimated from the Wilson interval, which remains accurate for fractions near 0 and 1~\citep{Brown2001}.
\section{Results and Verification} \label{sec::result}

\begin{deluxetable}{ccc}
	\centering 
	\tabletypesize{\footnotesize} 
	\tablecaption{Total number of BCGs and Those hosting AGN in Each Redshift Bin
	\label{table::agn_number}}
	\tablecolumns{7} 
	\tablehead{ \colhead{Redshift Bin} & \colhead{AGN-hosting BCG} & \colhead{All BCG}}
	\startdata \vspace{0.0cm}
	$0.00-0.39$ & 0 & 141 \\
	$0.39-0.55$ & 3 & 142\\
	$0.55-0.71$ & 7 & 143\\
	$0.71-1.30$ & 12 & 137
	\enddata \vspace{-0.1cm}
\end{deluxetable}

\begin{figure}[ht]
	\begin{center}
		\includegraphics[width=0.9\columnwidth]{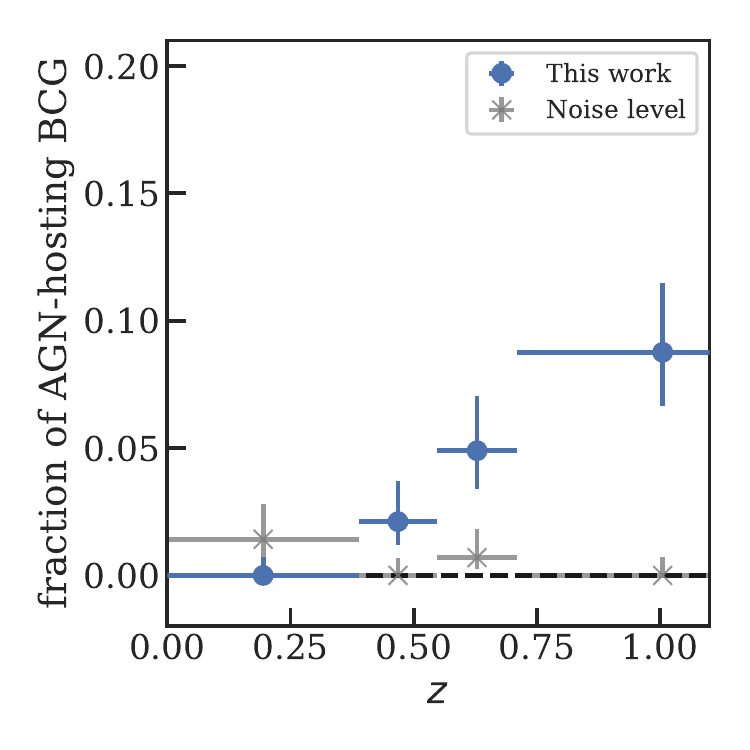}
		\caption{The fractions of AGN-hosting BCGs as a function of redshift. Blue points show the fractions from this work, which comes from the $W1-W2$ color residual in the SPT sample. The size of the error bar takes the binomial uncertainty into account. Gray crosses show the fraction of points that have residual less than -0.2, which we consider a ``noise level". This figure demonstrates that the fraction of AGN-hosting BCG increases with redshift.}
		\label{fig::frac_agn}
	\end{center}
\end{figure}
Table~\ref{table::agn_number} shows the number of AGN-hosting BCG in the four redshift bin, whereas the blue points in Fig.~\ref{fig::frac_agn} show the fraction of AGN-hosting BCGs in the four bins with their corresponding 68\% confidence intervals. We observe that the fraction is increasing with redshift in the SPT sample. Gray crosses show the fraction of points that have residuals less than -0.2. If the scatter is truly random, there should be a similar number of points below -0.2, compared to those above +0.2, which are classified as AGNs. This result implies that the increasing trend of the AGN-hosting BCGs fraction is not a result of the data quality. Such a trend has been suggested and shown in previous works~\citep[e.g.,][]{Hlavacek-Larrondo2013, McDonald2016, Birzan2017,Mo2020,Gupta2020}. In particular, we show that the fraction is $\sim$2\% at $z\approx0.5$ which is consistent with what~\citet{Somboonpanyakul2021a} found from looking at extreme central BCGs in clusters. We note that since some AGN-dominated galaxies will have poor photometric redshift constraints as they are estimated from the stellar spectrum and not AGN's power-law spectrum, we might misidentify these galaxies in our BCG finding algorithm. This implies that the number of BCGs with central AGNs found in this work gives a lower limit on the AGN-hosting BCG fraction, and the actual evolution could be even stronger. 

One way to evaluate the observed result with the chosen bins is to perform Fisher's Exact Test in order to see whether there are any associations between two categorical variables. The result shows that we can reject the null hypothesis of independence with P-value = 0.00045, meaning that there is a statistically significant association ($>99.9\%$) between redshift and whether or not BCGs host AGNs.

\begin{figure*}[ht]
	\begin{center}
		\includegraphics[width=1.95\columnwidth]{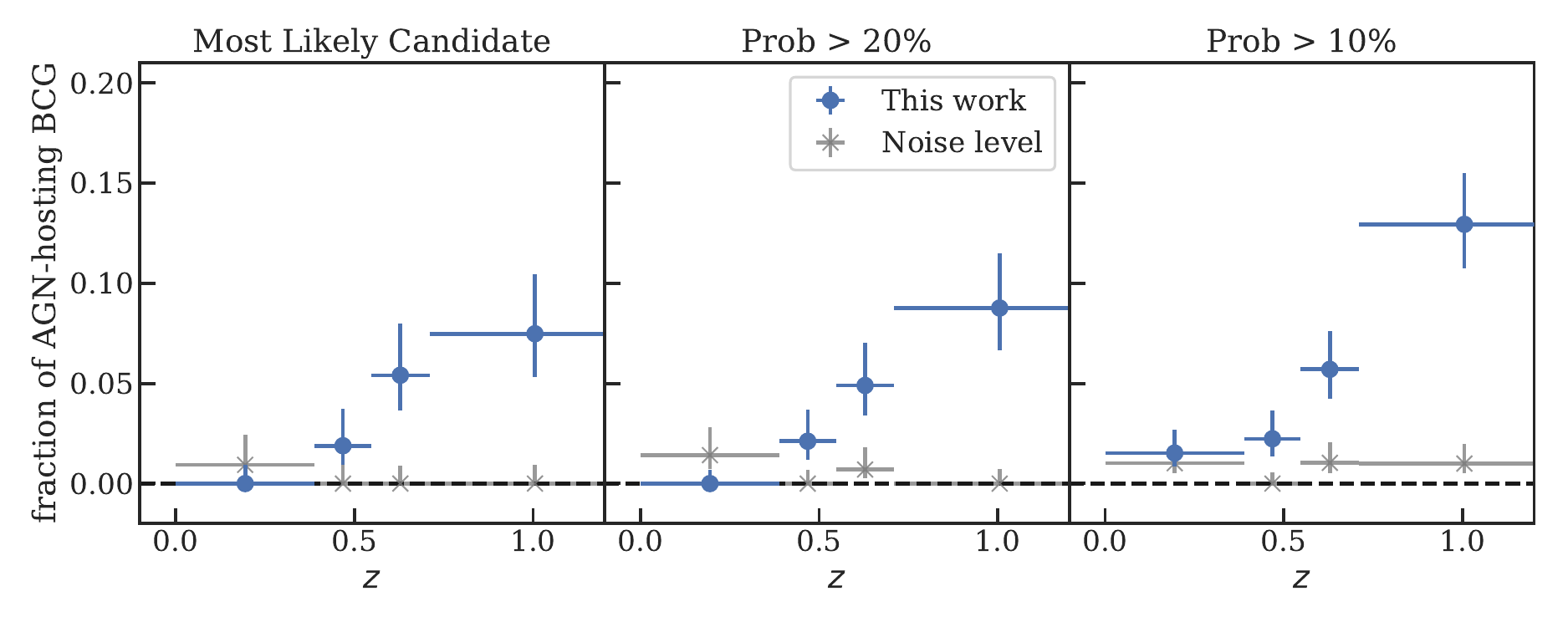}
		\caption{All three panels show the fractions of AGN-hosting BCGs, similar to Fig.~\ref{fig::frac_agn}. The left panel (triangles) shows the fraction when we consider only the most likely BCG for each cluster. The middle panel (dots) includes all possible BCGs with probability higher than 20\% while the right panel (squares) includes those with probability higher than 10\%. With all scenarios, the fraction remains increasing as a function of redshift.} 
		\label{fig::frac_agn_sampling}
	\end{center}
\end{figure*}

The approaches taken in this work are: (i) we include all BCG candidates with probability higher than 20\% in our sample, instead of picking only one BCG per cluster, and (ii) we calculate the fraction of BCGs with AGNs over the total number of BCG candidates, and not the fraction of clusters with AGN-hosting BCGs over the total number of clusters. The reason for these two assumptions is that we want to include AGN-hosting BCGs from systems with more than one obvious BCG, which are typical for merging systems such as the Coma cluster~\citep{Zwicky1933}, and the Bullet cluster~\citep{Markevitch2004}. We perform consistency checks to address both of these assumptions. Fig.~\ref{fig::frac_agn_sampling} shows the fraction of AGN-hosting BCGs when we consider the most likely BCG candidates, every BCG candidate with the probability higher than 20\%, and every BCG candidate with the probability higher than 10\%, respectively. This figure shows that the increasing trend of the fraction of AGN-hosting BCGs over redshift remains consistent in all three scenarios, regardless of how we select BCGs. 

\begin{figure}[ht]
	\begin{center}
		\includegraphics[width=0.98\columnwidth]{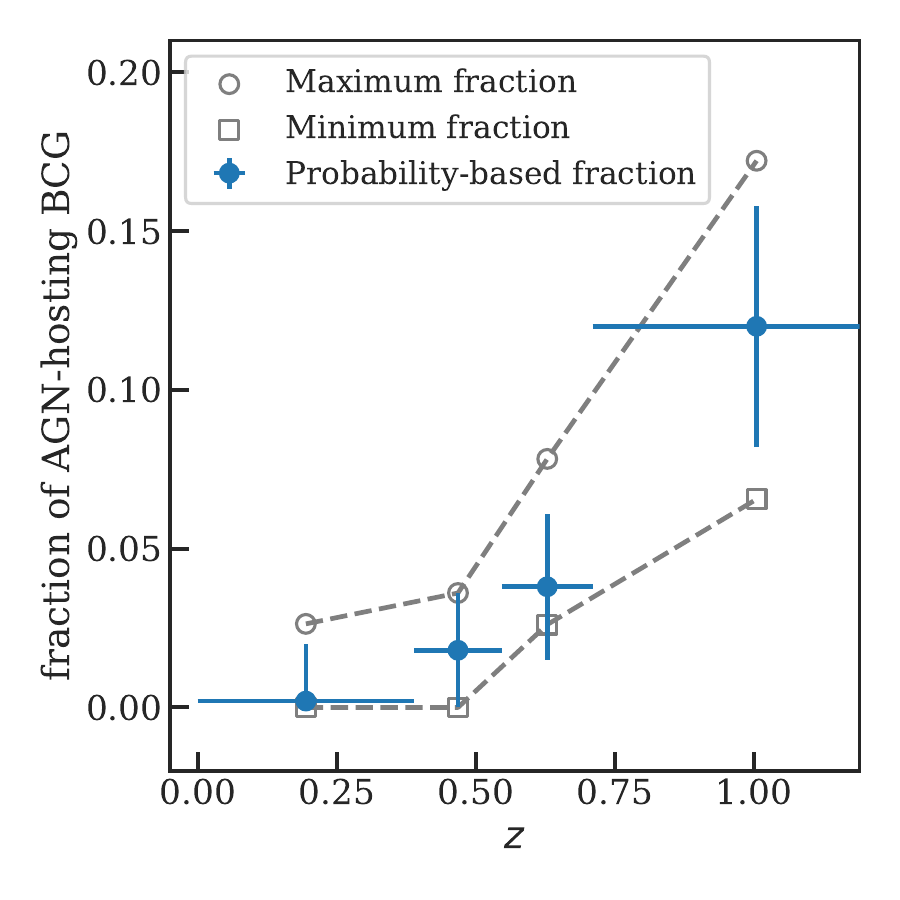}
		\caption{The fractions of clusters with AGN-hosting BCGs, similar to Fig.~\ref{fig::frac_agn}. The blue points assume one BCG per cluster and incorporate the probability of being a BCG, estimated from Section~\ref{sec::bcg}. The empty gray dots are for clusters which have any of their potential BCGs to be considered as AGNs, which is equivalent to the maximum fraction. The empty gray squares only include clusters whose BCG candidates are all considered AGNs, which is the minimum fraction. This figure shows that all of these definitions qualitatively give the same conclusion to our initial results.}
		\label{fig::frac_agn_method}
	\end{center}
\end{figure}

On the other hand, Fig.~\ref{fig::frac_agn_method} shows the results when we use different definitions of AGN fractions. The gray and blue points in Fig.~\ref{fig::frac_agn_method} are calculated with the total number of clusters as a denominator, instead of the number of BCG candidates. For the blue points, we consider one BCG per cluster and include both the probability of being BCGs, as calculated in Section~\ref{sec::bcg} and the uncertainty of the mid-IR color for each BCG to emphasize the fact that the uncertainties of identifying BCGs and BCG colors are higher at high redshift. The empty gray dots are the largest possible fractions, which are calculated from clusters that have any of their potential BCGs to be considered as AGNs, while the empty gray squares are the smallest possible fractions by counting only clusters which have all of their BCG candidates to be classified as AGNs. This figure illustrates that all of these definitions qualitatively give the same conclusion to our initial results.

\begin{figure*}[ht]
	\begin{center}
		\includegraphics[width=2.0\columnwidth]{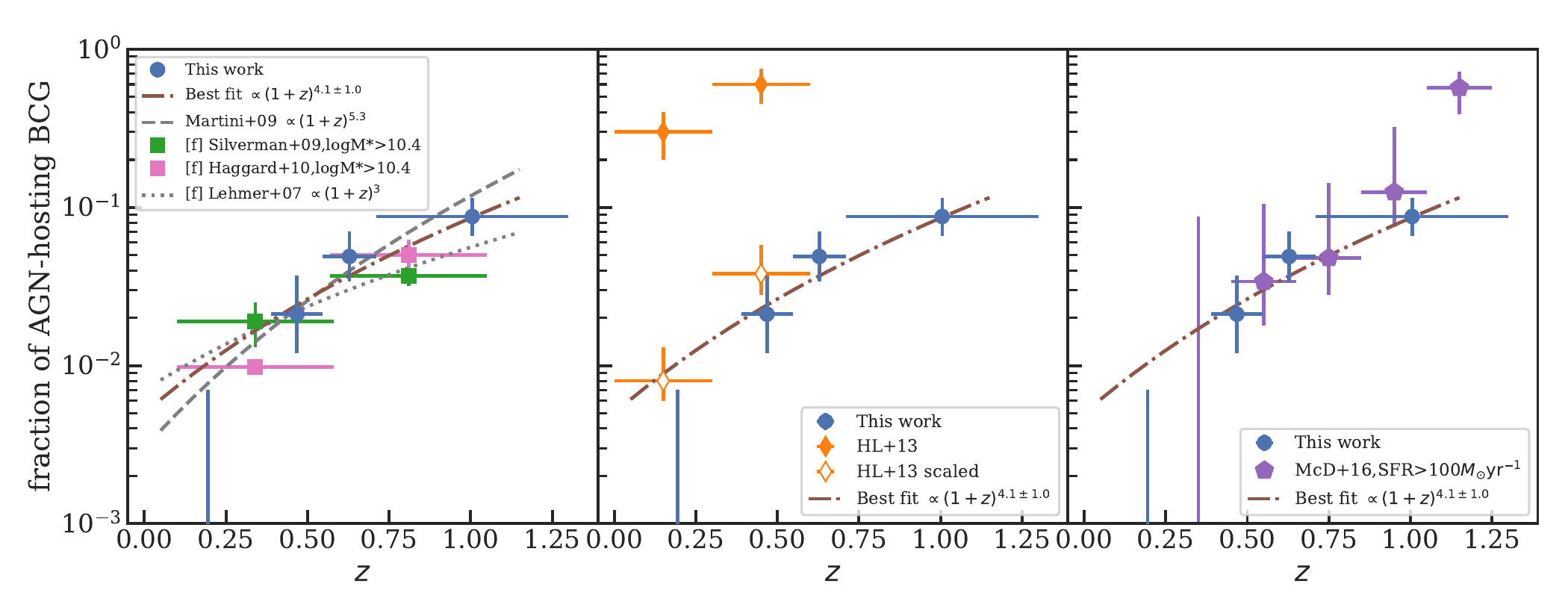}
		\caption{Comparison of the fraction of clusters with AGN-hosting BCGs to other published works. 
			Left: Green and pink squares are the field AGN fraction from~\citet{Silverman2009} and~\citet{Haggard2010}, respectively. The gray dotted line indicates pure luminosity evolution $\propto(1+z)^{3}$, suggested by~\citet{Lehmer2007}, while the gray dashed line shows the $(1+z)^{5.3}$ line, suggested by~\citet{Martini2009}. The brown dash-dotted line displays the best-fit model in the form $(1+z)^{4.1\pm1.0}$.
			Middle: Orange diamonds show the estimated fractions of BCGs with bright AGNs ($\rm{log}_{10}(L_{X,{\rm nuc}} \,[\rm{erg\, s^{-1}}])>42.2$) from~\citet{Hlavacek-Larrondo2013} with orange open diamonds showing the scaled version by changing the total number of clusters to a larger parent sample which~\citet{Hlavacek-Larrondo2013} had drawn from since this work only focuses on highly X-ray-luminous clusters ($L_{X}>3\times10^{44}\,{\rm erg/s}$).
			Right: Purple points show the fraction of BCGs with SFR $>100\,M_{\odot}\,\rm{yr}^{-1}$ from~\citet{McDonald2016}.}
		\label{fig::frac_agn_comp}
	\end{center}
\end{figure*}

We compare our results with the AGN fraction in field galaxies to determine whether there is a difference in the fractions between the two environments. The green and pink squares in the left panel of Fig.~\ref{fig::frac_agn_comp} show the field X-ray AGN fractions from the zCOSMOS survey~\citep{Silverman2009} and the \textit{Chandra} Multiwavelength Project results~\citep[ChaMP;][]{Haggard2010}, respectively. The results from our work are consistent with these two results, suggesting that the source of fuel required for AGN accretion in field galaxies could be similar to that in the brightest cluster galaxies. Additional evidence for the AGN fraction evolution in field galaxies has been seen in other works. For example,~\citet{Lehmer2007} finds an evolution in early type galaxies ($z\sim0.7$) consistent with the $(1+z)^3$ pure luminosity evolution model. The gray dotted line in Fig.~\ref{fig::frac_agn_comp} shows the curve for $(1+z)^3$ although it is only intended to be illustrative since it is arbitrarily normalized. The dashed line instead shows the curve for $(1+z)^{5.3}$. This line is first suggested by~\citet{Martini2009} who show the AGN fraction of cluster members to increase as $\sim(1+z)^{5.3}$ for AGN above an X-ray luminosity $L_x>10^{43}\,\rm{erg\,s^{-1}}$, hosted by luminous galaxies. We also fit the power law model ($\propto(1+z)^{\alpha}$) to the blue points in Fig.~\ref{fig::frac_agn_comp} and find a power law exponent $\alpha=4.1\pm1.0$, as shown in the brown dash-dotted line, which is consistent with the results from both~\citet{Lehmer2007} and~\citet{Martini2009}. Nevertheless, there are caveats regarding the relationship between cluster BCGs and field galaxies. One concern is that the AGN selection criteria for both BCGs and field galaxies are different, making it difficult to make a direct comparison between the two. In addition, according to the work about the evolution of AGN luminosity which shows that AGNs in galaxies tend to be brighter at high redshift~\citep{Silverman2008,Hasinger2005}, we would naturally expect to find a higher AGN fraction at high redshift since we usually selected AGNs based on a certain luminosity threshold.

\section{Discussion} \label{sec::discussion}
The results obtained in Section~\ref{sec::result} demonstrate that given a single-burst stellar population model, there is an increase in the fraction of AGN-hosting BCGs with redshift. This finding is consistent with previously published studies~\citep[e.g.,][]{Hlavacek-Larrondo2013, Birzan2017}, which focus on different samples with distinct selection effects. In particular, the results from~\citet{Hlavacek-Larrondo2013} and ours, as shown in the middle panel of Figure~\ref{fig::frac_agn_comp}, show the same trend of increasing fraction of AGN-hosting BCGs with redshift. However, the normalizations are vastly different. \citet{Hlavacek-Larrondo2013} claimed that the fraction of active BCGs is 30\% at $z\approx0.1$ and 60\% at $z\approx0.5$. On the other hand, we show that the fraction is less than 20\% at all redshift bins. A possible explanation is the very different samples that these works consider. While this study is based on an effectively mass-selected sample of clusters, with no consideration of X-ray properties, the sample used by \citet{Hlavacek-Larrondo2013} focuses solely on highly X-ray-luminous clusters ($L_{X,{\rm cluster}}>3\times10^{44}\,\rm{erg\,s^{-1}}$), which show clear X-ray cavities. To put these two studies on the same scale, we modify the denominator used by \citet{Hlavacek-Larrondo2013} in calculating the AGN fraction to account for the full parent population of clusters from which their sample of 32 clusters was drawn. Given that this previous work included a subsample of clusters drawn from the REFLEX~\citep{Bohringer2004}, eBCS~\citep{Ebeling2000}, MACS~\citep{Ebeling2001}, and SPT-XVP~\citep{McDonald2013} surveys, we consider these surveys in their entirety as the total population (the denominator) when calculating the AGN fraction. With this rescaling, we find that the fraction of BCGs hosting powerful AGN is consistent between our work and \citet{Hlavacek-Larrondo2013}, and that the observed evolution is consistent between both studies. We note that AGN are identified in very different ways between these two samples, but they are both selecting at the extreme end -- the few most mechanically powerful and the most IR-bright outbursts. The fact that these two works agree after the aforementioned rescaling is applied is reassuring, though a much more thorough analysis (and proper bias correction) is needed before conclusions about the co-evolution of jet power and mid-IR emission can be made.

An increase of the fraction of BCGs hosting central AGNs with redshift suggests that the accretion rates of the supermassive black holes in the BCGs are higher at high redshift since AGN luminosity is proportional to accretion rate. Several works about the relation between the mean black hole accretion rates and the cavity (kinetic)/quasar (radiative) power of the central AGN~\citep{Churazov2005,Russell2013} have shown that as black hole accretion increases in the BCGs, the cavity power of the AGN also increases to counteract the cooling from the accretion in a form of a negative feedback cycle. However, as the black hole accretion rates rise to near the Eddington limit, the cavity/jet power seems to be saturated, and the radiative power tends to dominate at this level of accretion. The fact that the radiative power from AGNs usually promotes more cooling in the ICM, instead of preventing it, suggests that a well-regulated feedback system between a central black hole and its host cluster is no longer possible at high accretion, implying that some galaxies might not have a fully established AGN feedback loop at this redshift range.

A similar conclusion has been reached from the work related to star forming galaxies~\citep{Webb2015,McDonald2016,Bonaventura2017}, which have shown that the fraction of starburst BCGs is higher at high redshift ($z>1$). Specifically, \citet{McDonald2016} found the fraction of BCGs with SFR over $10\,M_{\odot}\,\rm{yr}^{-1}$ to be $34\pm5\%$ at $0.25<z<1.25$, compared to $\sim$1--5\% at $z\sim0$. The right panel of Figure~\ref{fig::frac_agn_comp} compares the fraction of AGN-hosting BCGs in this work with the fraction of starburst BCGs (${\rm SFR_{\rm BCG}}>100\,M_{\odot}\,\rm{yr}^{-1}$) from \citet{McDonald2016}, demonstrating that in the center of cluster environments both massive starburst galaxies and bright AGNs behave similarly. These two results strongly hint that AGN feedback might not be as effective to prevent overcooling at high redshift as we have previously thought. 

All of these results lead us to suspect that the reason for the observed redshift trend and the breakdown of AGN feedback at high redshift comes from the fact that there is an abundance of cold gas at that redshift. Typically in the local universe, BCGs grows by merging with gas-poor satellites without triggering any AGN activity. However, BCGs at high redshift could grow by merging, instead, with gas-rich members. Cold gas from the mergers could be a source of fuel for increasing AGN activities in the center of clusters. This is consistent with the picture we get from the studies of starburst BCGs~\citep{Webb2015,McDonald2016} since cold dense clouds from gas-rich mergers could provide enough matter required for creating new stars. Further evidence supporting a gas-rich merger explanation includes the prevalence of cluster galaxies with massive CO or cold gas reservoirs at high redshift~\citep{Noble2017,Noble2019,Hayashi2018,Markov2020} and the detections of molecular gas in many BCGs~\citep{Dunne2021}. This scenario can also explain recent studies about the cool core (CC) fraction which show no sign of evolution over the same redshift range~\citep{McDonald2017,Ruppin2020}. If AGN feedback breaks down at high redshift, one would expect that the CC fractions of clusters would be higher since more gas should have been cooled near the center. However, if black hole accretion and star formation in high-redshift BCGs are fueled by something other than cooling of the hot gas, such as gas-rich mergers~\citep{Barnes1991,Hopkins2006}, it would be reasonable to think that the trends of AGN- and starburst-hosting BCG fractions would be different from the trend of CC fraction. If this observed increase in AGN activity is linked to gas-rich mergers, rather than ICM cooling, we would expect to see an increase in scatter in the $P_{cav}$ vs $L_{cool}$ relation~\citep{Rafferty2006} at $z>1$. 

Another possible scenario to explain the trend of high AGN-hosting BCG fraction at high redshift has to do with cluster mergers. It has been shown both in simulations~\citep{Fakhouri2010} and observations~\citep{McDonald2017} that the cluster merger rate is significantly higher at high redshift. Major mergers between two clusters have the potential to disrupt a tightly regulated AGN feedback loop and promote black hole accretion and star formation by potentially increasing the local turbulence of the system. This is consistent with the turbulent picture in the precipitation model for AGN feedback, called ``chaotic cold accretion (CCA)'', which states that turbulence is a key component to drive nonlinear thermal instability and extended condensation~\citep{Voit2015,Gaspari2020}. Turbulent forcing can help stimulate precipitation and condensation by raising the velocity dispersion of the ambient medium, resulting in more black hole accretion and star formation~\citep{Voit2018}. With a more energetic environment in the early universe, it is reasonable to assume that the turbulence will be higher at high redshift, resulting in higher black hole accretion rates. The recent discovery of CHIPS 1911+4455, a merging galaxy cluster with a massive starburst in the center, provides strong evidence that mergers can indeed increase star formation~\citep{Somboonpanyakul2021a,Somboonpanyakul2021b}. With the development of the next generation X-ray observatories, such as Athena and Lynx, we will be able to directly measure motions in the hot gas and determine whether mergers of groups/clusters can boost cooling via an increase in turbulence.

Lastly, Fig.~\ref{fig::spitzer} shows that the BCG of the Phoenix cluster remains the most extreme AGN in the entire SPT-SZ sample, which is over a 2500 deg$^2$ area and spans all redshifts. In combination of the recent work from the CHiPS survey~\citep{Somboonpanyakul2021a}, which have confirmed that the Phoenix cluster hosts the most extreme BCG with the strongest cool core at $z<0.7$, the runaway cooling phase, as we have seen in the Phoenix cluster, is indeed extremely rare.

\section{Conclusion} \label{sec::conclusion}
In this work, we present results on the mid-IR colors of BCGs in SPT-selected galaxy clusters at $0 < z < 1.3$. This study allows us to track the evolution of BCG properties over $\sim$9 Gyr of cluster growth. In particular, we focus our work on black hole accretion in BCGs, which turns these central galaxies into bright AGNs. Our findings are summarized as follows: 

\begin{enumerate}
	\item Assuming a single-burst stellar population model, we find statistically significant evidence ($>99.9\%$) for a mid-IR excess in high-redshift BCGs compared to low redshift BCGs, suggesting an increase with redshift in the fraction of AGN-hosting BCGs in galaxy clusters over $0 < z < 1.3$. For the lower redshift bins ($z<0.6$), an increase is not statistically significant, and the results are compatible with the noise level. On the other hand, we see an increase in the fraction of BCGs with AGNs at high redshift bins ($z>0.6$), similar to what others have found in previous works~\citep{Hlavacek-Larrondo2013,McDonald2016,Birzan2017}.
	
	\item We show that our results are consistent with both the evolution of the fraction of AGNs in field galaxies~\citep{Silverman2009,Haggard2010} and the fraction of starburst BCGs~\citep{Webb2015,McDonald2016}, suggesting that the reason for the evolution of both AGN and starburst fraction could come from the fact that more cold gas is available in the early universe. This should lead to a higher level of gas-rich mergers in BCGs, which could fuel both AGN activity and star formation in the center of clusters. There remain some caveats about the direct comparison between cluster and field galaxies ranging from selection criteria to the evolution of AGN luminosity.
	
	\item Another possible explanation for the increase in the fraction of AGN-hosting BCGs with redshift could be a higher level of local turbulence from dynamically active galaxy clusters at high redshift, leading to elevated cooling and subsequent black hole accretion. However, for this scenario it is difficult to explain the similarity to the trends in the field galaxies.
	
	\item We do not see any additional cluster with a BCG that is as extreme in the mid-IR color as the Phoenix cluster. In other words, the Phoenix cluster likely hosts the most extreme central AGN in the SPT sample.
\end{enumerate}

An enhancement of AGN activity in BCGs at high redshift compared to low redshift, which is similar in magnitude to the increase observed in field galaxies and cluster members, suggests that this increased AGN activity is not related to cooling flows, but rather to accretion of gas-rich satellites at early times. Such accretion events ought to throw off the precise cooling/feedback balance in the centers of clusters -- responsible for preventing runaway cooling flows -- leading to a less tightly regulated feedback loop at early times. Further studies with deeper and higher angular resolution mid-IR imaging, such as the upcoming \textit{James Webb Space Telescope}~\citep[JWST;][]{Gardner2006}, will be required to better understand the evolution of AGN feedback and its impact on galaxy clusters.

\acknowledgments
Acknowledgments.\\
\textit{Facilities}: Pan-STARRS, Chandra X-ray Observatory (ACIS), Hubble Space Telescope (ACS/WFC), the Nordic Optical Telescope (ALFOSC) \\

\textit{Software:} astropy~\citep{Astropy2018}, CIAO~\citep{Fruscione2006}, pandas~\citep{McKinney2010}, seaborn~\citep{Waskom2016}

T.\ S.\ and M.\ M.\ acknowledge support from the Kavli Research Investment Fund at MIT, and from NASA through Chandra grant GO5-16143.

F.R. acknowledges financial supports provided by NASA through SAO Award Number SV2-82023 issued by the Chandra X-Ray Observatory Center, which is operated by the Smithsonian Astrophysical Observatory for and on behalf of NASA under contract NAS8-03060.

This publication makes use of data products from the Wide-field Infrared Survey Explorer, which is a joint project of the University of California, Los Angeles, and the Jet Propulsion Laboratory/California Institute of Technology, funded by the National Aeronautics and Space Administration.

The South Pole Telescope is supported by the National Science Foundation through grant PLR-1248097. Partial support is also provided by the NSF Physics Frontier Center grant PHY-1125897 to the Kavli Institute of Cosmological Physics
at the University of Chicago, the Kavli Foundation, and the Gordon and Betty Moore Foundation grant GBMF 947.

Funding for the DES Projects has been provided by the U.S. Department of Energy, the U.S. National Science Foundation, the Ministry of Science and Education of Spain, the Science and Technology Facilities Council of the United Kingdom, the Higher Education Funding Council for England, the National Center for Supercomputing Applications at the University of Illinois at Urbana–Champaign, the Kavli Institute of Cosmological Physics at the University of Chicago, the Center for Cosmology and Astro-Particle Physics at the Ohio State University, the Mitchell Institute for Fundamental Physics and Astronomy at Texas A\&M University, Financiadora de Estudos e Projetos, Fundação Carlos Chagas Filho de Amparo à Pesquisa do Estado do Rio de Janeiro, Conselho Nacional de Desenvolvimento Científico e Tecnológico and the Ministério da Ciência, Tecnologia e Inovação, the Deutsche Forschungsgemeinschaft and the Collaborating Institutions in the Dark Energy Survey.

The Collaborating Institutions are Argonne National Laboratory, the University of California at Santa Cruz, the University of Cambridge, Centro de Investigaciones Enérgeticas, Medioambientales y Tecnológicas–Madrid, the University of Chicago, University College London, the DES-Brazil Consortium, the University of Edinburgh, the Eidgenössische Technische Hochschule (ETH) Zürich, Fermi National Accelerator Laboratory, the University of Illinois at Urbana-Champaign, the Institut de Ciències de l’Espai (IEEC/CSIC), the Institut de Física d’Altes Energies, Lawrence Berkeley National Laboratory, the Ludwig-Maximilians Universität München and the associated Excellence Cluster Universe, the University of Michigan, the National Optical Astronomy Observatory, the University of Nottingham, The Ohio State University, the OzDES Membership Consortium, the University of Pennsylvania, the University of Portsmouth, SLAC National Accelerator Laboratory, Stanford University, the University of Sussex, and Texas A\&M University.

\bibliographystyle{yahapj}


\end{document}